\documentclass[11pt]{article}

\usepackage{amsthm}
\usepackage{amsmath}
\usepackage{amssymb}
\usepackage{natbib}
\usepackage{bm}
\usepackage[dvips]{graphicx}
\usepackage{longtable}
\usepackage{fullpage}
\usepackage{setspace}
 \usepackage{soul}
 \usepackage{multirow}

\newcommand{\iid}  {\stackrel{\rm iid}{\sim}}
\renewcommand{\a}{\alpha}

\newcommand{\kv}{{\bf k}}

\begin{document}
\title{Bayesian Estimation of Negative Binomial Parameters\\ with Applications to RNA-Seq Data}
\author{Luis  Le\'on-Novelo\thanks{To whom correspondence should be addressed: 
Department of Biostatistics,
UTHealth School of Public Health, 
P.O. Box 20186 , Houston, TX 77225. Email:  \texttt{luis.g.leonnovelo@uth.tmc.edu}}\\
UT Health Science Center
\and
Claudio Fuentes\\
Oregon State University
\and
Sarah Emerson\\
Oregon State University
}

\date{}
\maketitle
\begin{abstract}
RNA-Seq data characteristically exhibits large variances, which need to be appropriately accounted for in the model.  We first explore the effects of this variability on the maximum likelihood estimator (MLE) of the overdispersion parameter of the negative binomial distribution, and propose instead the use an estimator obtained via maximization of the marginal likelihood in a conjugate Bayesian framework. We show, via simulation studies, that the marginal MLE can better control this variation and produce a more stable and reliable estimator.  We then formulate a conjugate Bayesian hierarchical model, in which the estimate of overdispersion is a marginalized estimate and use this estimator to propose a Bayesian test to detect differentially expressed genes with RNA-Seq data. We use numerical studies to show that our much simpler approach is competitive with other negative binomial based procedures, and we use a real data set to illustrate the implementation and flexibility of the procedure.

\end{abstract}

\bigskip
{\it Keywords}: Bayes estimation; Hierarchical models; Maximum likelihood estimation; Hypothesis testing; Negative binomial; RNA-Seq analysis; DEG.

\newpage
\section{Introduction}

The modeling of RNA-Seq data has proved to be problematic: the underlying mechanism resembles some kind of Bernoulli experiment, but the actual data typically exhibit overdispersion and other features that often prevent using simple models (such as the binomial) to perform the analysis. In this context, a nice introduction to the RNA-Seq technology and the statistical analysis leading to a Poisson model is given in \cite{salzman2011statistical}. 

Alternatives to the Poisson model have also been considered, including a normal approximation to the binomial \citep{Feng2010} and the beta-binomial generalized linear model \citep{zhou2011beta} among others.  However, the overdispersion problem still persists, motivating researchers to develop new approaches that can explicitly address this issue, such as the overdispersed Poisson model \citep{Auer:Doerge:2010} and the negative binomial model \citep{Robinson:Smyth:2008,Anders:Huber:2010, di2011nbp, graze2012allelic}, which we consider in this paper. 

The negative binomial density  can be parametrized as 
\begin{equation}\label{NB_density_alpha}
p(k\mid \alpha,\mu)=
\frac{1}{{\cal B}(k+1,\a^{-1}) (k+\a^{-1})}
\left(\frac{\alpha\mu}{\alpha\mu+1}\right)^k
\left(\frac{1}{\alpha\mu+1}\right)^{1/\alpha}
,\quad k=0,1,\dots,
\end{equation}
where ${\cal B}(x,y)=\Gamma(x)\Gamma(y)/\Gamma(x+y)$ is the beta function and $\mu,\alpha\geq 0$. Under this parameterization $E(k)=\mu$ and $Var(k)=\mu+\alpha\mu^2$, and therefore the overdispersion is explicitly determined by the value of the parameter $\alpha$. Despite this natural advantage, the negative binomial model still presents some challenges. For instance, maximum likelihood estimation can be problematic, as it shares some of the same features with the binomial model. \cite{olkinm:petkau:videk1981} observed that two similar samples from a binomial distribution with success probability $p$ and sample size $n$ (both unknown), can lead to very different maximum likelihood estimators of the sample size.  To address this problem, \cite{carroll:lombard:1985} proposed estimating $n$ by  first integrating the likelihood  with respect to a beta distribution on $p$, and then obtaining an estimate of $n$ by maximizing this integrated likelihood. Applying the same idea to the negative-binomial distribution in the estimation of the overdispersion parameter $\alpha$, it can be found that such marginalization improves the properties of the estimates.  This result has a direct application in the estimation of the expression levels from RNA-Seq data, as the negative binomial distribution has become popular in the modeling of such data. 


\subsection{Models for RNA-Seq data}\label{sec:RNA-Seq}


The general problem consists of detecting differentially expressed genes (DEG) in RNA-Seq data sets coming from two different populations, from now on referred as to \emph{treatments}. The data can be represented in a table whose entry $(i,j)$ corresponds to the RNA read counts aligning to gene $i$ in individual $j$. The set of reads originating from individual $j$ (the column of the table)  is referred as to the \emph{library} for individual $j$. The \emph{library size} or abundance $s_j$ for individual $j$, is a positive number proportional to the coverage or sampling depth of library $j$ and can be proportional to the total number of reads in (or the sum over) column $j$. However, this sum can be sometimes dominated by the counts of few highly expressed genes. In addition,  the read counts depend not only on the underlying mean expression level but also on the library size and therefore, any statistical analysis must appropriately account for (or adjust for) the difference in library sizes. 

In this context, \cite{Robinson:Smyth:2008} propose a test for DEG  when only a few read counts per treatment are observed. 
Their method assumes that the read counts are negative binomial and that the overdispersion parameter $\alpha$ is the same across the genes. More importantly, they propose a conditional test (similar in nature to the Fisher exact test) to detect  whether the counts associated with a given gene follow a negative binomial distribution with the same parameters under both treatments. 
The test assumes that $\alpha$ is known and conditions on the sum of all the counts (over both treatments) for gene $i$, considering different library sizes.  \cite{Robinson01112007} \citep[see also][]{chenetal:2014} relax the assumption of constant overdispersion and, for each gene, 
estimate a different $\alpha_i$ using an empirical Bayes procedure. In both cases, the estimated values of the overdispersion parameters are taken as the true value. The bioconductor package  \textit{edgeR} by \cite{Robinson01012010} implements both approaches.

\cite{Anders:Huber:2010} propose an alternative solution in their procedure DESeq. They also cosider a negative binomial distribution for the gene counts, but  assume that genes with similar mean counts have similar variances.  That is, if $\mu_i\approx \mu_j$, then $\a_i\approx \a_j$, where $\mu_i$ and $\a_i$ are the mean and overdispersion parameters of the negative binomial distribution for the counts of gene $i$. Specifically, they model the variances as a smooth function of the means and propose an unbiased estimator for the variance $\nu_i$ performing a local regression of $\nu_i$ on $\mu_i$. The adjusted $\nu_i$'s are treated as the variance true values.  Finally, for each gene, they test whether the counts have the same mean across the  two different treatments. In order to select differentially expressed genes, DESeq builds on the procedure proposed by \cite{Robinson:Smyth:2008}.

Instead, \cite{Auer:Doerge:2010} discuss the design of experiments targeting the identification of DEG using RNA-Seq data. They emphasize the importance of replication (having more than one individual per treatment), randomization and blocking.  For each gene separately, they  model the log-mean of the read counts as the sum of the log-library size and the treatment effect. That is, for gene $i$, they take $\log \mu_i=\log s_j+\tau_{i\ell}$, where $\tau_{i\ell}$ is the treatment effect for $\ell=1,2$.  In their approach, the library size $s_j$ for  individual $j$ represents its overall number of reads and is estimated before fitting the model and treated as known.  The treatment effect can be estimated through a Poisson regression with ``exposures'' $s_j$. Using the estimated means, they suggest a statistical test for DEG that considers the possible overdispersion of the data (with respect to the Poisson model), based on  the likelihood ratio test for the Poisson regression. Auer and Doerge (A\&D) also suggest a balanced block design (BBD) to control for lane and batch effects, and then include both batch and lane effects in the linear model for the log-mean. In order to make the means identifiable, in addition to biological replication, it is necessary to split or ``bar code'' each individual into the number of lanes of the sequencing machine. The same statistical test can be applied under the BBD. 

A different method is considered by \cite{hardcastle:kelly:2010} in their baySeq procedure. They also start from a negative binomial model, but then take a Bayesian approach. However, rather than using a Bayesian prior distribution for $\alpha$, they construct an estimated, data-dependent prior, using a combination of empirical Bayes, maximum likelihood, and quasi-likelihood. 
The baySeq method is implemented in the bioconductor package of the same name.

Alternatively \cite{wu2013new} propose a negative binomial Bayesian model with a log-normal prior for the overdispersion  parameter $\alpha$. Every gene is assumed to follow this model and the parameters of the log-normal distribution are estimated using an empirical Bayes procedure. Then, for every gene $i$, the model is applied a point estimate of  the overdispersion parameter $\alpha$ is obtained. Finally, they use these point estimates and the Wald test to determine if a gene is differentially expressed. More recently, \cite{law2014voom} proposed  a log transformation of the read counts after adjusting for library size, and to analyze the transformed data as the intensity continuos data arising from a microarray experiment, but without assuming equal variances across genes. Instead, they propose using a modified version of the empirical Bayes procedure available in the limma package \citep{smyth2005limma} that incorporates the mean-variance trend as part of the statistical model. They discuss two alternatives to model the mean-variance trend, \emph{limma-trend} and \emph{voom}.

\cite{rapaport2013comprehensive} compares the performance of DESeq, \textit{edge} R, baySeq, and other methods (but not A\&D)
using benchmarks data sets. They conclude that there is no a uniform best method in all the scenarios of their comparisons. However, they remark that apparently the methods based on the negative binomial distribution (mentioned above) perform  better in terms of specificity and sensitivity.

Here, we propose a Bayesian model based inference procedure to detect differentially expressed genes (DEG) in two populations. 
Our approach allows for libraries of different sizes and 
assumes a different $\alpha_i$  for each gene. It declares a  gene to be DEG based on its posterior probability of being differentially expressed. In contrast to \cite{Anders:Huber:2010}, our model does not assume any functional relationship between $\alpha_i$ and $\mu_i$ and in contrast to \cite{hardcastle:kelly:2010}, we specify the prior distributions for $\mu$ and $\alpha$, avoiding the possible overfitting that sometimes results from data-dependent priors. On the other hand, we propose an empirical Bayes procedure to estimate hyper-parameters  of the prior distribution of $\alpha$, but contrary to  \cite{wu2013new}  that implements a Wald test, we propose a Bayesian hypothesis framework using Bayes factors comparing both the null and alternative hypotheses in order to make decisions. 
Simulation results show that our approach competes favorable with the inference obtained by using DESeq, baySeq and A\&D.

The rest of the paper is organized as follows: in Section \ref{sec:overdispersion} we discuss the estimation of the overdispersion parameter, and in Section \ref{sec:testing} we introduce the Bayesian framework to test for DEG. In Section \ref{sec:simulate} we examine the performance of our procedure under a number of simulation scenarios, and compare our approach to DESeq, A\&D, baySeq and edgeR. In Section \ref{sec:application}  we illustrate  our method analyzing a real data set and contrast our results to those obtained by other methods. We end with a short discussion of the method in Section \ref{sec:disc} and include an appendix with some of the technical details.

\section{Estimation of the Overdispersion Parameter}\label{sec:overdispersion}

 \begin{table}[b]
 \caption{\label{tab:smallnb} \textsc{\small Five negative binomial samples each with maximum likelihood estimates for the mean $\mu$ and overdispersion parameter $\alpha$.  The reported standard deviation is $\sqrt{\hat \mu+ \hat \alpha {\hat \mu}^2}$ }.}
 \begin{center}
 \begin{tabular}{c|c|c|c}
 Sample&$\hat{\alpha}$&$\hat \mu$&Std. Dev.\\
 \hline
 \vspace{-.1in}&&\\
 $2 ,3 ,4 ,5 ,8$&$0.000$&$4.40$& $2.097$\\
$2 ,3, 4, 5, 9$&$0.052$&$4.60$&$2.386$\\
$2,  3, 4,  3, 11$&$0.210$&$4.60$&$3.006$\\
$2,  3, 4,  2, 12$&$0.329$&$4.60$&$3.402$\\
\hline
\end{tabular}
\end{center}
\end{table}
When only a small sample from a negative-binomial distribution is available (a common feature in the analysis of RNA-Seq data) the maximum likelihood estimator (MLE) $\hat{\alpha}$, of the overdispersion parameter is unstable.  
For instance, 
when  the sample mean is greater than the sample variance,  the maximum is attained at a negative value, and in such cases the estimator is truncated at zero. If this is not the case, the MLE of the overdispersion parameter typically exhibit a large variability. Table \ref {tab:smallnb} illustrates the effect on the estimation of $\alpha$ of small  perturbations in a sample of size five. We observe that even small changes in a single observation of the sample may greatly affect the estimation of the overdispersion parameter and consequently have an effect on the inference about $\mu$, increasing of the standard error of up to $60\%$, without a significant increment in the estimated value of the mean.

\begin{figure}\label{fig:boot}
\begin{center}
\includegraphics[width=3.5in,height=3in, angle=270 ]{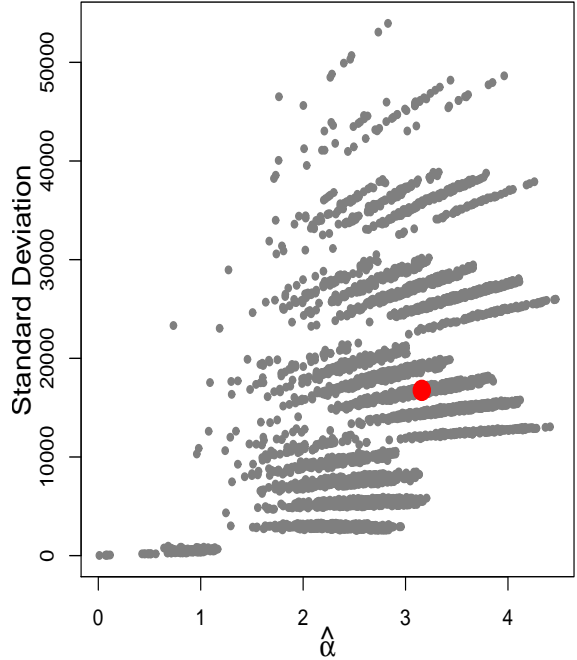}\\
\end{center}
\caption{\label{fig:boot}\textsc{\small Values of $\hat \alpha$ ($x$-axis) and the corresponding standard deviation of $\hat \mu$ ($y$-axis) for $5000$ bootstrap samples of the RNA-Seq data shown in the bacteroides counts example.  The minimum standard deviation is $20.53$ and the maximum is $53949.37$. The red dot indicates the value for the original sample.} }
\end{figure}

In a more realistic scenario, consider the following data from a pilot project to determine bacterial community differences across time for individuals at high risk for type 1 diabetes.  Using 454 pyrosequencing of amplified 16S rRNA genes, the counts of bacteroides from ten children are 118, 131, 136, 176, 274, 1022, 1675, 14137, 15714 and 60886.
In order to illustrate the variability in the MLE's of $\mu$ and $\alpha$, we took $5000$ bootstrap samples of these data. The estimated values of $\alpha$ and the corresponding estimated standard deviation of $\hat{\mu}$ obtained for each sample are depicted in  Figure \ref{fig:boot}.  We observe a huge bootstrap variability suggesting that the sampling variability will be of equal size.

The problems with the MLE of the overdispersion parameter  have been discussed in the literature and \cite{Robinson:Smyth:2008} introduce a quasilikelihood estimator that performs better than the MLE when estimating $\alpha$. Here, we propose that the variability of the estimator can be controlled using a marginal MLE instead. This marginal MLE of $\alpha$ is obtained by first integrating the parameter $\mu$ out of the likelihood  in (\ref{NB_density_alpha}) with respect to a conjugate distribution and then by maximizing  the resulting marginal likelihood with respect to $\alpha$.

To accomplish  this, we first observe that the $\text{F}$-distribution is a conjugate prior for $\mu$ in this setting and assume that $\mu \vert \alpha$ $\sim$ $\text{F}(\nu_1,\nu_2)$, where $\nu_1$ and $\nu_2$ denote the degrees of freedom.  In order to obtain a simple closed form expression for the model and, at the same time, preserve its hierarchical structure, we take $\nu_1=2a_\mu$ and $\nu_2=2a_\mu/\alpha$, with $a_\mu>0$. This way,  the conditional density is
\begin{equation}\label{cond_density}
p(\mu\mid\a)=\frac{1}{{\cal B}(a_\mu,a_\mu/\a)} \frac{\a^{a_\mu} \mu^{a_\mu-1}}{(1+\a\mu)^{a_\mu+a_\mu/\a}},
\end{equation}
and for $\kv= (k_1,\dots,k_J)$, the integrated likelihood is given by
\begin{eqnarray}\label{neg_bin_mu_marginalized}
p(\kv\mid\alpha)&=&\int_0^\infty p(\kv\mid\alpha,\mu)p(\mu\mid\alpha)\,d\mu\\
&=&\left[ \prod_j \frac{1}{{\cal B}(k_{j}+1,\a^{-1}) (k_{j}+\a^{-1})    }\right]
\frac{{\cal B}(a_\mu+\sum_j k_j,(J+a_\mu)/\alpha)}{{\cal B}(a_\mu,a_\mu/\alpha)},\nonumber
\end{eqnarray}
where  $p(\kv\mid\alpha,\mu)=\prod_j p(k_j \mid\alpha,\mu)$ is the likelihood function and the $k_j$'s are iid $\text{NegBin}(\mu,\a)$, for all $j$.
Finally, to complete the specification of the model, we have to choose a value for the constant $a_\mu$. From (\ref{cond_density}) we observe that $E(\mu\mid\alpha)=a_\mu/(a_\mu-\alpha)$, when $a_\mu>\alpha$ and the expectation does not exist when $a_\mu\leq \alpha$. Therefore, in oder to reduce the impact that the parameter $\alpha$ has on the random behavior of $\mu$, we prefer small values of $a_\mu$ that produce densities less informative with heavier tails.

It follows that the three estimators under consideration are:
\begin{enumerate}
\item  $\hat{\alpha} = \mbox{argmax}_\a \{ \max_\mu p(\kv\mid\alpha,\mu)\}$, the MLE, 
\item  $\tilde{\alpha} = \mbox{argmax} \; p(\kv\mid\alpha)$, the marginal MLE, and
\item $\hat{\alpha}_{QL}$, the quasilikelihood estimator proposed by  \cite{Robinson:Smyth:2008}.
\end{enumerate}
We compare the performance of each one of these estimators in different scenarios and summarize the results in Table \ref{table:simresults_Imu}. 
We observe that for small sample sizes the MLE $\hat \alpha$ does not perform well: it is often truncated to zero and has mean squared error much larger than the marginal MLE $\tilde \alpha$, which by construction, is always greater than zero.  Although the MLE improves as $\mu$ or $n$ increases, it is still outperformed by the marginal MLE.   Furthermore, we observe that in practically all the considered settings, the marginal MLE produce estimators  closer (on average) to the true value and better, in terms of the mean square error (MSE), than the quasilikelihood estimator, offering a more stable and reliable estimator.

\begin{table}
\caption{\label{table:simresults_Imu} \textsc{\small 
Comparison of estimators of the overdispersion parameter of the negative binomial distribution.  
The MLE is $\hat{\alpha}$, the marginal MLE is $\tilde{\alpha}$, and 
the quasilikelihood is $\hat{\alpha}_{QS}$.
Each row represents a different negative binomial distribution with mean $\mu$, overdispersion $\alpha$, and sample size $n$.  
 The values of the mean, median, and mean squared error (MSE) were obtained based on $1000$ negative binomial samples.
The last column gives the proportion of times that the solution to the likelihood equations was negative, which truncates the MLE to zero.  
 In the calculations $a_\mu$ was set equal to 0.01.}}
\begin{center}
{\small
\begin{tabular}{lll| lll| lll| lll| l}
 \multicolumn{3}{c}{True Values}&
  \multicolumn{3}{c}{$\tilde{\a}$}&
\multicolumn{3}{c}{$\hat{\a}$}&
\multicolumn{3}{c}{$\hat{\a}_{QS}$}
\\
$\mu $ &$\alpha$ &$n$& mean &median & MSE &mean &median & MSE &mean &median & MSE & $\% \hat{\a}_{\max}<0$\\
\hline
\multicolumn{13}{c}{Coefficient of variation, $\sqrt{1/\mu+\alpha}$ equal to 1.1 }\\
1&0.21&3&0.101&0&0.15&6.408&0&608.101&1.841&0.605&11.525&70.5 \\ 
 1&0.21&10&0.187&0&0.161&0.315&0&0.416&0.467&0.317&0.311&55.1 \\  
 1&0.21&50&0.195&0.126&0.046&0.217&0.149&0.053&0.379&0.359&0.083&27.5 \\  
 10&1.11&3&0.918&0.607&1.305&1.33&0.6&31.503&1.893&1.197&4.748&13.1 \\  
10&1.11&10&1.032&0.928&0.381&1.049&0.934&0.447&1.41&1.291&0.722&0\\ 
 10&1.11&50&1.089&1.071&0.058&1.092&1.073&0.059&1.339&1.324&0.146&0 \\  
100&1.2&3&0.9&0.701&0.775&0.895&0.645&0.988&1.578&1.074&2.934&1.3 \\ 
 100&1.2&10&1.112&1.05&0.24&1.114&1.045&0.261&1.473&1.352&0.612&0 \\  
 100&1.2&50&1.186&1.176&0.048&1.187&1.177&0.049&1.439&1.423&0.146&0 \\  
 1000&1.209&3&0.888&0.707&0.645&0.856&0.639&0.784&1.529&1.057&2.496&0.1 \\  
 1000&1.209&10&1.097&1.039&0.205&1.092&1.031&0.221&1.437&1.326&0.503&0 \\  
 1000&1.209&50&1.193&1.19&0.045&1.192&1.19&0.045&1.441&1.433&0.138&0 \\  
\hline
\multicolumn{13}{c}{Coefficient of variation equal to 1.5 }\\
1&1.25&3&0.286&0&1.478&14.838&0&1386.5&2.76&1.341&13.938&49.2 \\ 

 1&1.25&10&0.816&0.436&1.579&1.568&0.674&33.71&1.09&0.991&0.948&24.5 \\  
 1&1.25&50&1.162&1.087&0.364&1.238&1.151&0.394&1.106&1.102&0.158&0.6 \\  
 10&2.15&3&1.444&0.932&3.329&3.252&1.051&148.871&2.946&2.21&7.804&10.9 \\  
 10&2.15&10&1.986&1.76&1.153&2.09&1.834&1.429&2.616&2.448&1.508&0 \\  
 10&2.15&50&2.106&2.053&0.198&2.125&2.072&0.204&2.534&2.509&0.367&0 \\  
 100&2.24&3&1.78&1.266&2.99&1.97&1.281&4.882&3.278&2.316&9.028&0.9 \\  
 100&2.24&10&2.095&1.952&0.813&2.15&1.995&0.911&2.95&2.792&2.112&0 \\  
 100&2.24&50&2.212&2.186&0.151&2.224&2.198&0.154&2.844&2.82&0.631&0 \\  
 1000&2.249&3&1.709&1.309&2.491&1.818&1.304&3.949&3.173&2.333&8.563&0.3 \\ 
 1000&2.249&10&2.123&2.001&0.707&2.166&2.038&0.774&3.061&2.88&2.393&0 \\  
 1000&2.249&50&2.22&2.195&0.147&2.23&2.205&0.15&2.913&2.879&0.749&0 \\  
\end{tabular}
}
\end{center}
\end{table}

\section{Testing for  Differentially Expressed Genes}\label{sec:testing}

Suppose that we have two treatments and let $J_1$ and $J_2$ be the number of individuals under treatments 1 and 2 respectively. Assume that the individuals labeled with $j=1,\dots,J_1$ received the first treatment and the individuals labeled with $j=J_1+1,\dots,J_1+J_2$ received the second treatment. Denote by $k_{ij}$ the number of reads aligned to gene $i$ in individual $j$ and recall that  $s_j$ corresponds to the library size or abundance for the individual $j$. 
In this context, we observe that the problem of testing whether gene $i$ is differentially expressed can be looked as a model selection problem and consider a Bayesian framework to compare the hypotheses
\begin{equation}\label{twocompetingmodel}
\begin{array}{rl}
H_{0i}: &k_{ij}\sim \text{NegBin}(s_j\mu_{i0},\alpha_{i0}),   \quad j=1,\dots,J_1+J_2\\
&\mu_{i0}\mid\a_{i0}\sim\text{F}(2a_\mu,2a_\mu/\alpha_{i0})\hbox{  and }
\alpha_{i0}\sim \text{Gamma}(u_0,v_0)\\

H_{1i}:&k_{ij}\sim \text{NegBin}(s_j\mu_{i1},\alpha_{i1}),\quad j=1,\dots,J_1\\
&k_{ij}\sim \text{NegBin}(s_j\mu_{i2},\alpha_{i1}),\quad j=J_1+1,\dots,J_1+J_2\\
&\mu_{i1},\mu_{i2}\mid\a_{i1}\iid \text{F}(2a_\mu,2a_\mu/\alpha_{i1}) \hbox{ and } 
\alpha_{i1}\sim \text{Gamma}(u_1,v_1),
\end{array}
\end{equation}
where $\text{Gamma}(u,v)$ is the gamma distribution with expectation $u/v$. 
Observe that the means of the negative binomial distributions are adjusted by the library size $s_j$ (a typical feature of these models). In practice, the value of $s_j$ is assumed to be known or estimated form the data and taken as the true value. Here, we follow the strategy in \cite{Anders:Huber:2010}, and estimate estimate $s_j$ as 
\begin{equation}\label{libsize}
\hat{s}_j= \text{median}_i \frac{k_{ij}}{(\prod_{v=1}^{J_1+J_2} k_{iv})^{1/(J_1+J_2)}}.
\end{equation}

For the overdispersion parameters we consider only one value per gene under each hypothesis. Even though the model is flexible enough to allow for two different values of the overdispersion parameter $\alpha_{i1}$ under $H_{1i}$, in real applications the addition of an extra overdispersion parameter makes the estimation problem much more difficult and some times not feasible in situations when only a few counts are available. Nonetheless, notice that the hierarchical structure of this framework implicitly considers the marginal distribution of Section \ref{sec:overdispersion}, and therefore takes advantage of the stability properties of the marginal MLE of the overdispersion parameter. 

In order to complete the specification of the models,  we need to determine the values of the hyperparameters of the Gamma priors. Here we take an empirical Bayes approach and consider the following procedure: to determine $u_0$ and $v_0$ we estimate, for every gene $i$, the marginal MLE $\tilde{\alpha}_{i0}$ of the overdispersion parameter, assuming that the count means are the same in both treatments. Once the values of $\tilde{\alpha}_{i0}$'s are obtained,  we consider  $\tilde{\alpha}_{1,0},\dots,\tilde{\alpha}_{n_{genes},0}$ to be random sample from \text{gamma}$(u_0,v_0)$ and set $u_0$ and $v_0$ equal to  the MLE's. For the values of $u_1$ and $v_1$ we follow the same strategy, but using only the control group data. Of course, some variations can be considered, but in practice they lead to similar values for the hyperparameters and have little impact on the performance of the model.

Finally, to specify the prior probabilities $P(H_{0i})$ and $P(H_{1i})$ of the models we have a few alternatives. For instance, we can assume that the probabilities $P(H_{1i}), i=1,\dots,n_{genes}$ are observations from an underlying beta distribution and assign prior distributions for the beta distribution parameters and, therefore, induce a dependency in the posterior distribution across the genes. Instead, we can simply assume that  $P(H_{1i}):=\pi_1$ is the same for all genes, so the tests remain independent. If this is the case we observe that the ranks of the posterior probabilities do not change. That is, the list of $G$ genes with highest posterior probability using $P(H_{1i}):=\pi_1, \forall i,$  is the same as when using  $P(H_{1i}):=\pi_1^\prime,\forall i,$ with $0<\pi_1,\pi_1^\prime<1$. The choice of $\pi_1$ has an impact only on the estimation of the posterior expectation of the false discovery proportion (FDP) (see Section \ref{sec:simulate}). Hence, for each gene we can approximate the Bayes Factor (BF), in favor of the model $H_{1,i}$, $BF_{i,10}\equiv p(\kv_i\mid H_{1i})/p(\kv_i\mid H_{0i})$, and estimate $\pi_0=1-\pi_1$ following the strategy in \cite{wen2013robust}:
 \begin{enumerate}
 \item Sort the observed BFs in ascending order: $BF_{(1),10},\dots,BF_{(n_{genes}),10}$.
 \item Find $d_0= \max\{d: (1/d) \sum_{i=1}^d BF_{(i),10}<1\}$.
 \item Set $\tilde{\pi}_0=d_0/n_{genes}$ and $\tilde{\pi}_1=1-\tilde{\pi}_0$.
 \end{enumerate}
 
Observe that when the abundances $s_j$ are all equal to 1, it follows from \eqref{neg_bin_mu_marginalized} that 
the computation of $p(\kv_i\mid H_{mi})$, $m=0,1$, reduces to calculating (numerically) a one dimensional integral. When $s_j$ are not all equal to one we can compute the Bayes factor
$BF_{i,10}$ using the Gibbs sampling algorithm included in the Appendix \ref{app:gibbs}.


\section{Simulation Studies}\label{sec:simulate}

In this section we consider three different simulation studies to compare the performance of our method with DESeq,  baySeq, edgeR and the A\&D procedures:\begin{enumerate}
\item Study $1$:  Here we assume the abundances are known and equal to $1$, and only a small number of genes is considered.  
Two subscenarios are explored: (a) the values of the overdispersion parameter are set to make the simulation favorable for DESeq, and (b) the values of the overdispersion parameter are set to make the simulation not favorable for DESeq  
\item Study $2$: Here we assume the  abundances are unknown, and we use $10,000$ genes with $240$ differentially expressed.  We use the estimated means ($\mu$'s) from DESeq as the true values for the simulation. The overdispersion parameters  ($\alpha$'s) take values between 0 and 0.5 and are simulated from a $\text{Beta}(0.3,0.3)/2$ distribution.

\item Study $3$: Use the same conditions  as Study $2$, but we take the values of the overdispersion parameters $\alpha'$s equal to the DESeq estimate. 
\end{enumerate}
Observe that all three simulation studies are designed to meet the DESeq assumptions, with Studies $1$(a) and $3$ satisfying them exactly, while in studies $1$(b) and 2 the true value of the overdispersion parameter not a function of the mean. Table \ref{table:sim_scenarios} summarizes the simulation scenarios described above.  
 
\begin{table}
\caption{\label{table:sim_scenarios} \textsc{\small 
Summary of the simulation true parameters under the three scenarios described in the text. DESeq estimates means that the simulation true parameters of $\alpha$
were set equal to their DESeq estimated values as explained in the text.}}
\begin{center}
\begin{tabular}{ r|cc |c|c}
\multicolumn{1}{c}{Study}&\multicolumn{1}{c}{$N^o$ genes}&\multicolumn{1}{c}{$N^o$ DEG}
& \multicolumn{1}{c}{True $\alpha$'s}
&\multicolumn{1}{c}{Library Size}\\
\hline
1 a&\multirow{2}{*}{$10^2$}&\multirow{2}{*}{20}&\multicolumn{1}{c|}{DESeq estimates}&\multirow{2}{*}{all 1}\\
\cline{4-4}
1 b& & &\multicolumn{1}{c|}{Uniform(0,0.7)}\\
\hline
\multirow{2}{*}{2}&\multirow{3}{*}{$10^4$}&
\multirow{3}{*}{240}&\multirow{2}{*}{\text{Beta}(0.3,0.3)/2}&$s_1=s_3=0.8,$\\
& & & & $s_2=s_4=1$, and\\
\cline{4-4}
3& & &\multicolumn{1}{c}{DESeq estimates}&\multicolumn{1}{|c}{$s_3=s_6=1.2$}\\
\hline
\end{tabular}
\end{center}
\end{table}
In order to compare the performance of the procedures in terms of the false discovery proportion, we first define $d_i$ as an indicator variable to indicate the decision, that is $d_i=0$ when the $i$-th null hypothesis is accepted and $d_i=1$ otherwise. Likewise, we define the variable $r_i=0$ if indeed the null hypothesis $i$ in \eqref{twocompetingmodel} is true and $r_i=1$ if the alternative is true. This way, the FDP is defined as 
\begin{equation*}
FDP=\frac{\sum_i d_i(1-r_i)}{\sum_i d_i},
\end{equation*}
and the corresponding posterior expectation is
\begin{equation*}
{\rm E} (\mbox{FDP $\vert$ {\it data}}) = \frac{\sum_i d_i[1-P(H_{1i}\mid\kv_i)]}{\sum_i d_i}.
\end{equation*}
DESeq implements the \cite{Benjamini:Hochberg:1995} procedure to control the FDP.  In contrast, when testing the hypotheses in \eqref{twocompetingmodel}, we  declare the $n_{sel}$  genes with the highest $P(H_{1i}\mid \kv_i)$ 
as DEG and estimate the  corresponding expected FDP \citep[see][]{Newton:2004}. 

\begin{figure}[]
\caption{\label{fig:BayesFactor_100genes-A} \textsc{\small 
Comparison of different procedures to select DEG 
under the simulation scenario $1$(a).  The left panel compares the ROC curves, 
and the right panel depicts the simulation true FDP (solid) and the controlled FDP reported by the different approaches (dotted).  =
Scenario 1 Favors DESeq and baySeq,  $\a_i=\a_i^{DESeq}$.}The numbers in parentheses correspond to the Area Under the curves (AUC).}
\includegraphics[width=2.75in,height=3.25in,angle=270]{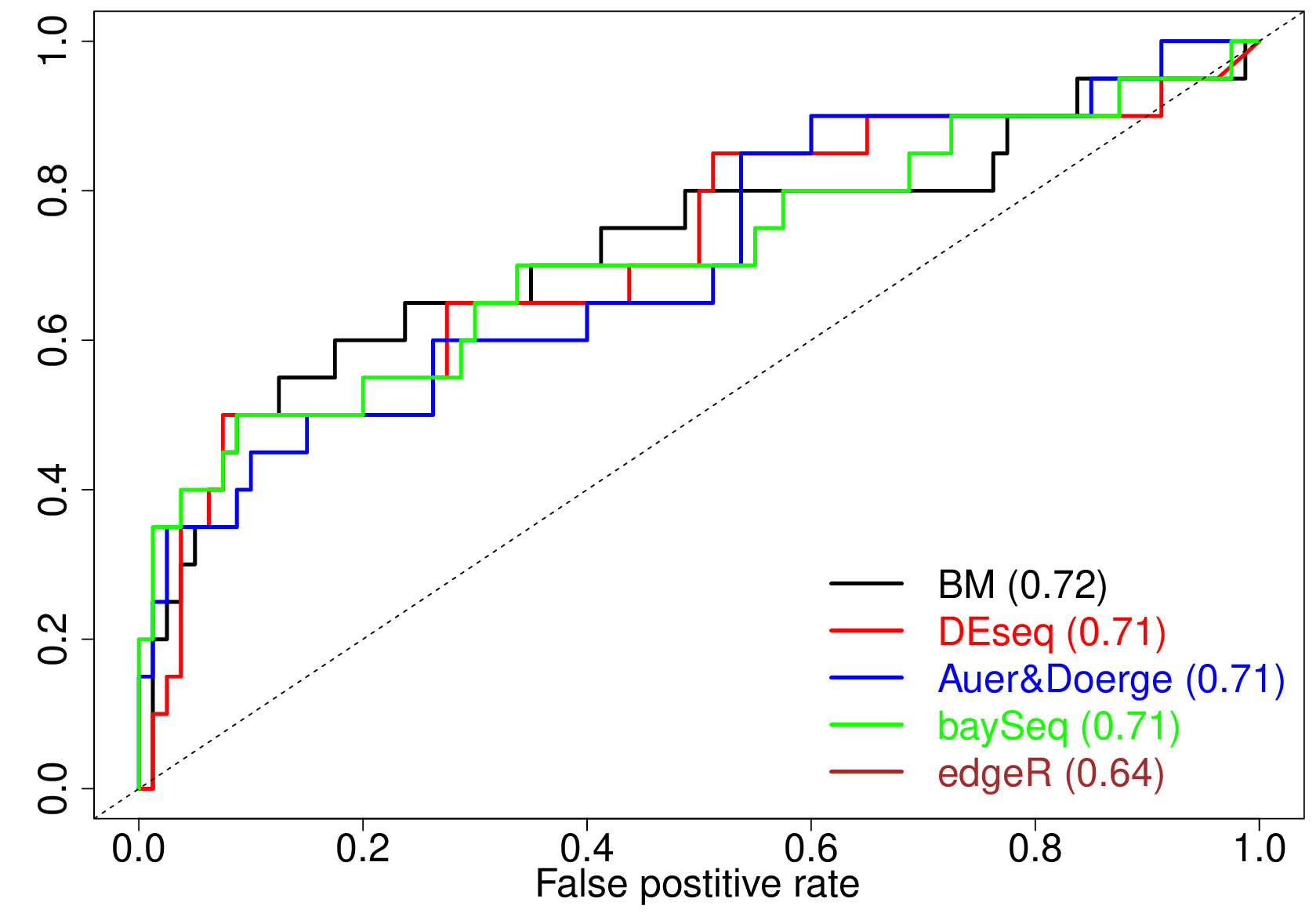}
\includegraphics[width=2.75in,height=3.25in,angle=270]{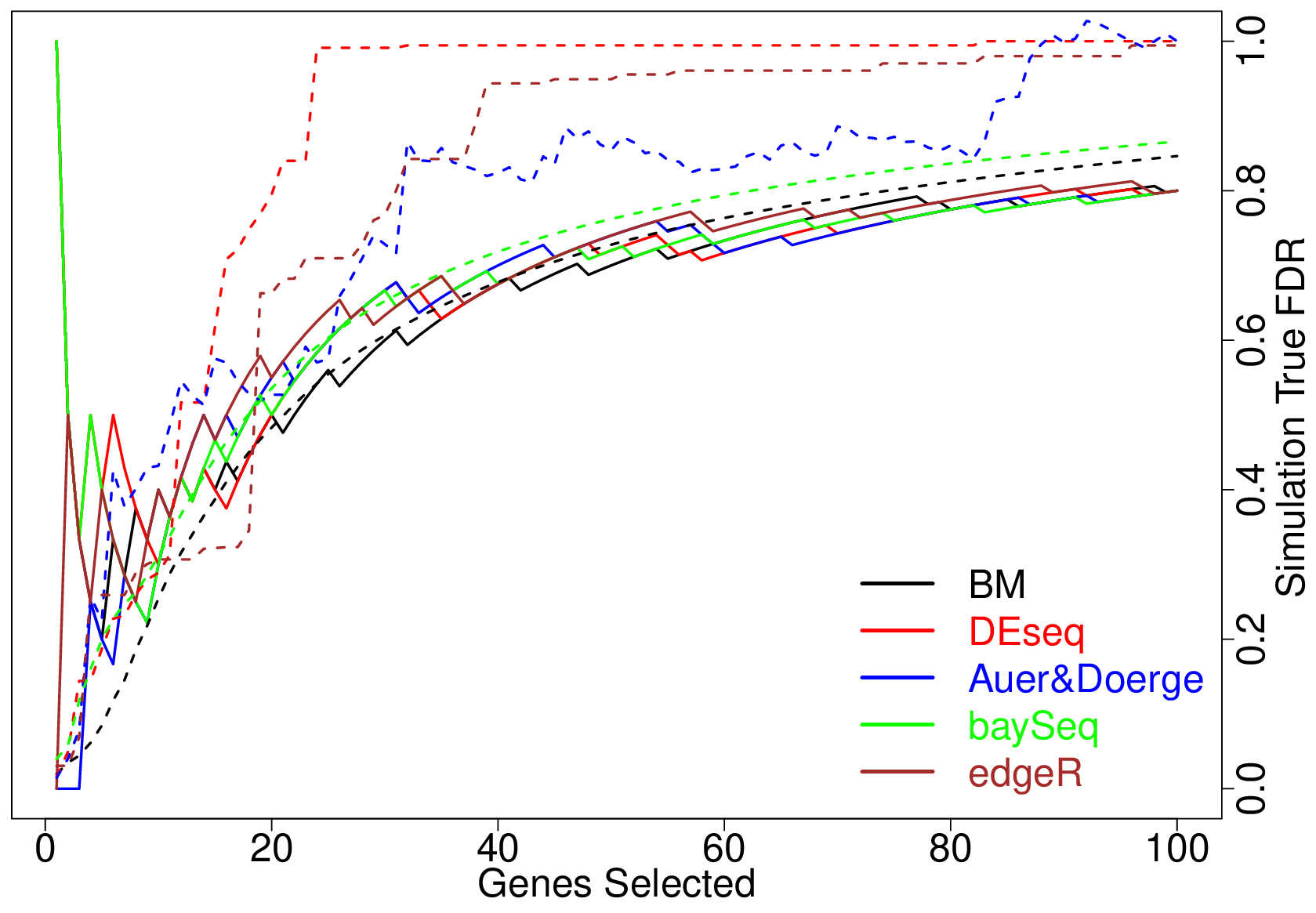}
\end{figure}

\begin{figure}[]
\caption{\label{fig:BayesFactor_100genes-B} \textsc{\small Comparison of different procedures to select DEG 
under the simulation scenario $1$(b).  The left panel compares the ROC curves, 
and the right panel depicts the simulation true FDP (solid) and the controlled FDP reported by the different approaches (dotted).    
In the simulation truth, $ \a_i\sim Unif(0,0.7)$. baySeq does not do well (left) and 
DESeq overestimates the FDP (right). The numbers in parentheses correspond to the Area Under the curves (AUC).
}}
\includegraphics[width=2.75in,height=3.25in,angle=270]{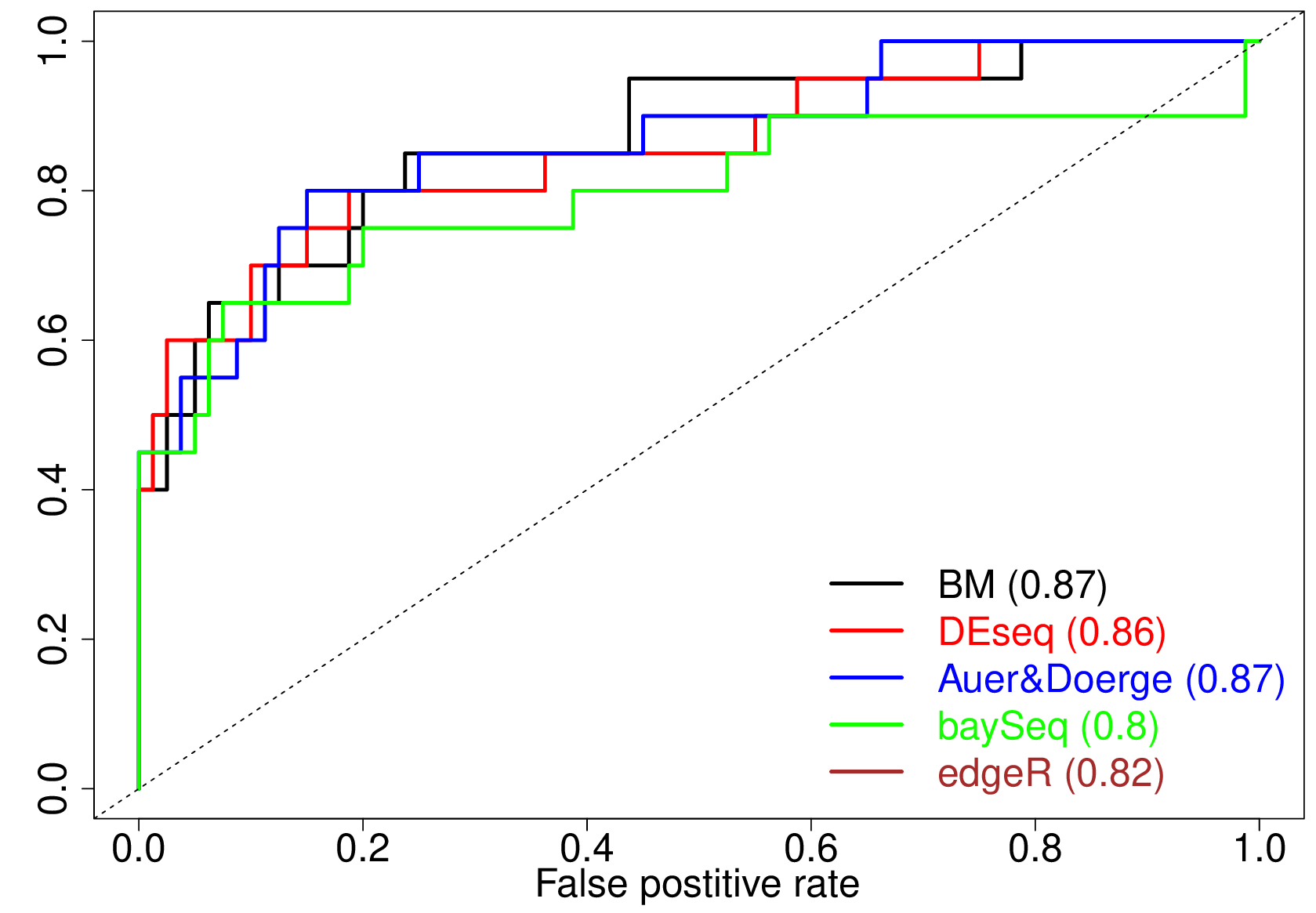}
\includegraphics[width=2.75in,height=3.25in,angle=270]{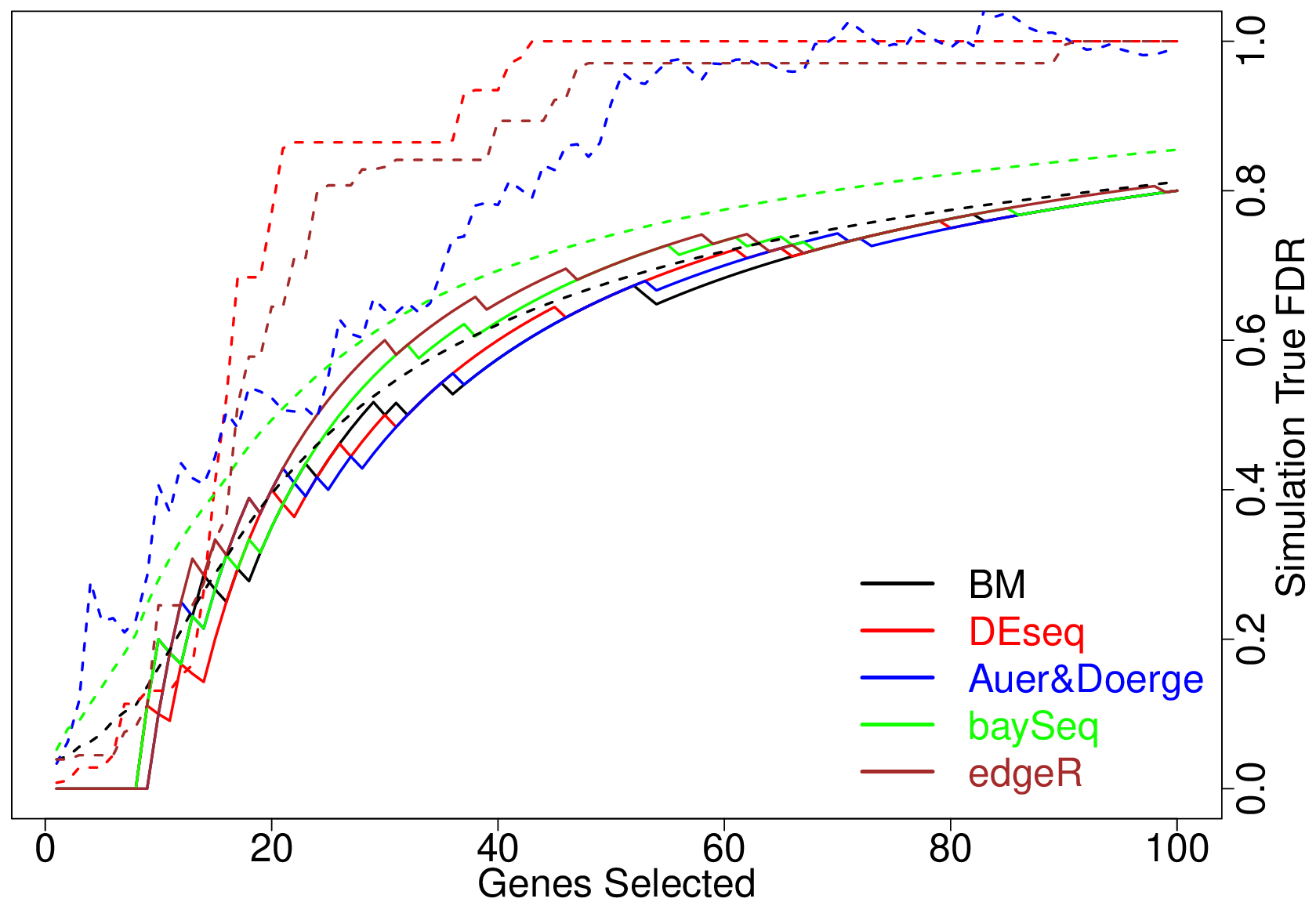}
\end{figure}

\subsection{Study 1:  Few genes and abundances equal to one}\label{sec:abundance-1}
We compare DESeq, baySeq, edgeR and A\&D to the proposed approach. 
In all analyses of simulated data in this paper, regardless of the approach,
we
used the simulation true abundances (as opposed to 
estimating the abundances in the way proposed by each approach). 
We simulate counts of  \emph{Treatment 1} with $J_1=3$ individuals  
as follows:
\begin{enumerate}
\item We used the data set ``TagSeqExample.tab'' included in the DESeq R library. More specifically, we used the 
first 100 genes whose estimated DESeq base means are greater than 10 in the 3 samples labeled with T1b,T2 and T3, and estimated the library sizes according to \eqref{libsize}.
\item We estimate $\mu_i^{DESeq}=\text{baseMean}$ and $\alpha_i$ as  $\alpha_{i}^{DESeq}=\text{fittedRawVar}/\text{baseMean}^2$.
The values \text{baseMean} and \text{fittedRawVar} are estimated by using the R function ``nbinomTest'' in the DESeq R library.
\item We used these estimates as the true values for the simulation in the \emph{Treatment 1} group. That is, we simulated $k_{ij}\iid \text{NegBin} (\mu_i,\a_i)$ for $j=1,2,3$ where $\mu_i$ and 
\begin{enumerate}
\item $\alpha_i$ are set to the estimated values in 2.  (Note that by using these values of $\alpha_i$ the simulation fully satisfies the DESeq 
assumptions, making this simulation the most favorable for DESeq.)
\item $\a_i\sim Unif(0,0.7)$ favoring our procedure.
\end{enumerate}
\item
For \emph{Treatment 2} we considered the same number of individuals, $J_2=3$. The first 20 genes are differentially expressed (they have different means from treatment 1)  while the remaining 80 share the same mean as their corresponding gene in the treatment 1 group. For all individuals we set the true abundances  for the simulation to be $s_j=1$ and assume them known in all analyses.  More specifically, $k_{ij}\iid \text{NegBin} (c_i \mu_i^{DESeq},\a_i^{DESeq})$ for $j=4,5,6$ with \
$c_1=6,c_2=5.75,c_3=5.5,\dots,c_{20}=1.25$.
This implies, for instance, that the true mean counts of the first DEG in the treatment 2 group is $6$-fold larger than its treatment 1 counterpart. 
\end{enumerate}

Figures \ref{fig:BayesFactor_100genes-A} and \ref{fig:BayesFactor_100genes-B}
compare the results of DESeq,  baySeq and A\&D with the proposed Bayes model (BM). We observe that even though the simulations settings are designed to favor the DESeq and baySeq assumptions, the Bayes model gives practically the same results in terms of ROC curves and true FDP. Furthermore, in these scenarios the BM procedure does a better job estimating and reporting the false discovery proportion (dotted line on the right panel in both figures) than the competitors.


\subsection{Studies $2$ and $3$:  Simulation study with abundances}\label{sec:study2-3}

Recall that DESeq assumes that genes with similar means have similar overdispersion parameters, that is, if $\mu_i\approx \mu_j$ then $\alpha_i\approx\a_j$. This condition is satisfied in Study 3, but not in Study 2. Specifically, we simulated a dataset with ten thousand genes and two treatments, with three individuals per treatment. The values for the simulation parameters were set based on the Drosophila data described in Section \ref{sec:application}. Then, we estimated the means of 9760 counts using DESeq and  compute the empirical deciles: $q_{0.1},\dots,q_{0.9}$ and the quantile $q_{0.95}$. Only the first 240 genes in the simulated data set are DEG. Under Study 2,
the $\alpha_1,\dots,\alpha_{10,000}\iid \text{Beta}(0.3,0.3)/2$.
  Under Study 3, for each gene, the $\alpha'$s are set according to the DESeq estimates. The simulated means of the 240 differentially expressed genes are shown schematically in Table \ref{table:means}. 

For the 9760 not differentially expressed genes, we used the means estimated by DESeq. 
For each gene $\alpha_i$ is the same in both treatments.
 Under Study 3, the $\alpha'$s  are set equal to their DESeq estimate. Finally, we estimate the abundances according to the algorithm in DESeq given in \eqref{libsize} and select the genes with highest posterior probabilities.

\begin{table}[]
\caption{\label{table:means} \textsc{\small Schematic representation of the simulated sample means. Here $q_p$ represents the
quantile $p$ of the 9760 genes selected at random from the Drosophila data of Section \ref{sec:application}.  The pattern of means given for the first $120$ genes is repeated in genes $121-240$.
Under Study 2, $\alpha_1,\dots,\alpha_{10,000}\iid \text{Beta}(0.3,0.3)/2$.
Under Study 3 the $\a'$ fulfill the DESeq assumptions explained in the text.}}
\begin{center}
\begin{tabular}{|c|cc|}
\multicolumn{1}{c}{Gene Index}&\multicolumn{2}{c}{Mean}\\
\hline
&&\vspace{-.1in}\\
&\multicolumn{1}{c}{Treatment 1}&Treatment 2\\
\hline
&&\vspace{-.1in}\\
$1-3$&$q_{0.10}$&$2q_{0.10}$\\
$4-6$&$q_{0.10}$&$q_{0.10}/2$\\
\hline
$7-9$&$q_{0.10}$&$3q_{0.10}$\\
$10-12$&$q_{0.10}$&$q_{0.10}/3$\\
\hline
$13-18$&$q_{0.20}$&$2q_{0.20}$\\
$19-24$&$q_{0.20}$&$q_{0.20}/2$\\
\hline
$25-96$&$\vdots$ &$\vdots$\\
\hline
$97-102$&$q_{0.95}$&$2q_{0.90}$\\
$103-108$&$q_{0.95}$&$q_{0.90}/2$\\
\hline
$109-114$&$q_{0.95}$&$3q_{0.95}$\\
$115-120$&$q_{0.95}$&$q_{0.95}/3$\\
\hline
\end{tabular}
\end{center}
\end{table}

\begin{figure}[t]
\caption{\label{fig:BMFvsDESeq_combo_Csimulationalpha_exp_v_50_p1nonrandom} \textsc{\small 
Comparison of ROC and FDP curves of different procedures under scenario 2 
($\alpha_i\iid\text{Beta}(0.3,0.3)/2$, $\tilde{\pi}_1=0.0275$ with simulation true $\pi_1=0.024$). The numbers in parentheses correspond to the Area Under the curves (AUC).
}}
\includegraphics[width=2.75in,height=3.25in,angle=270]{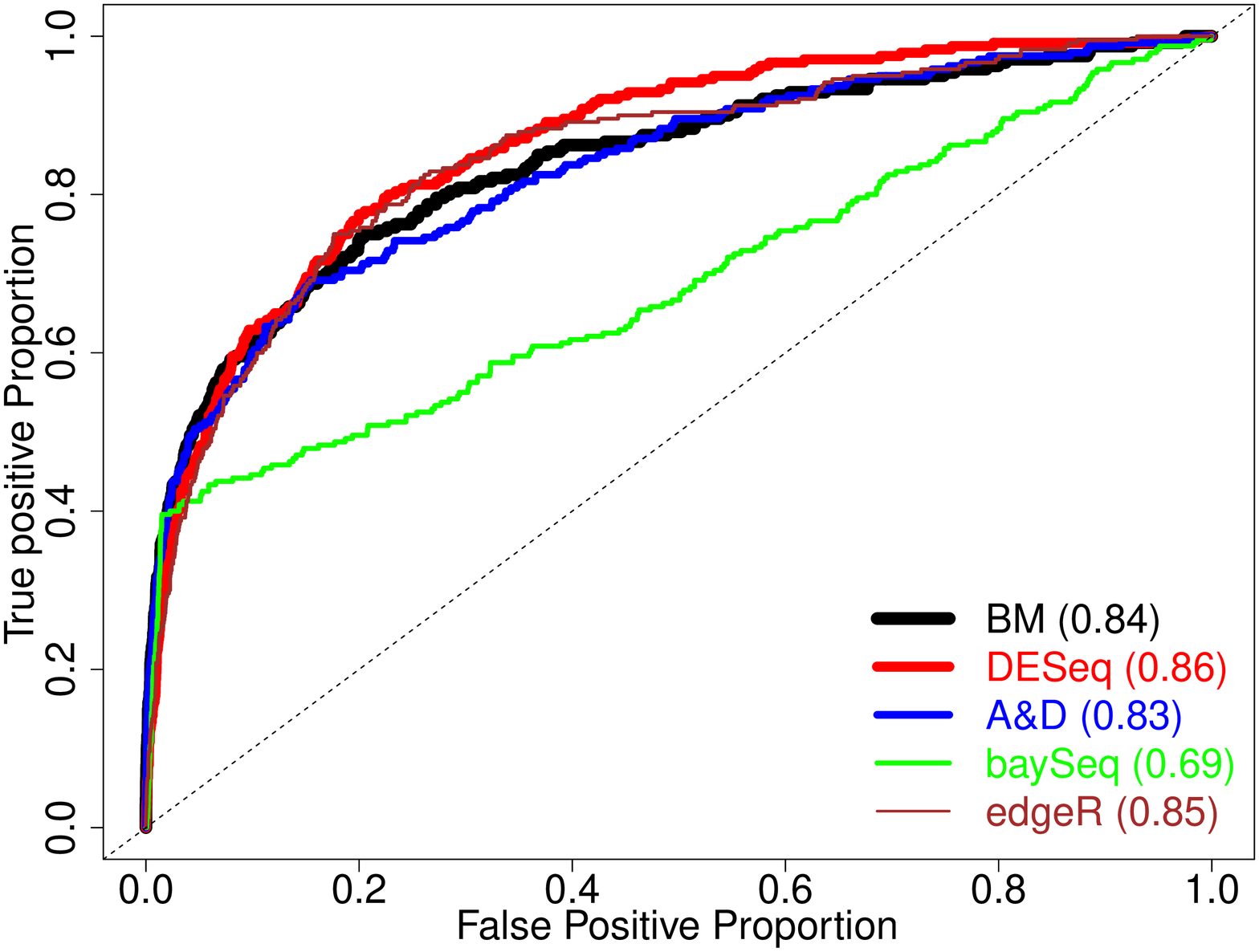}
\includegraphics[width=2.75in,height=3.25in,angle=270]{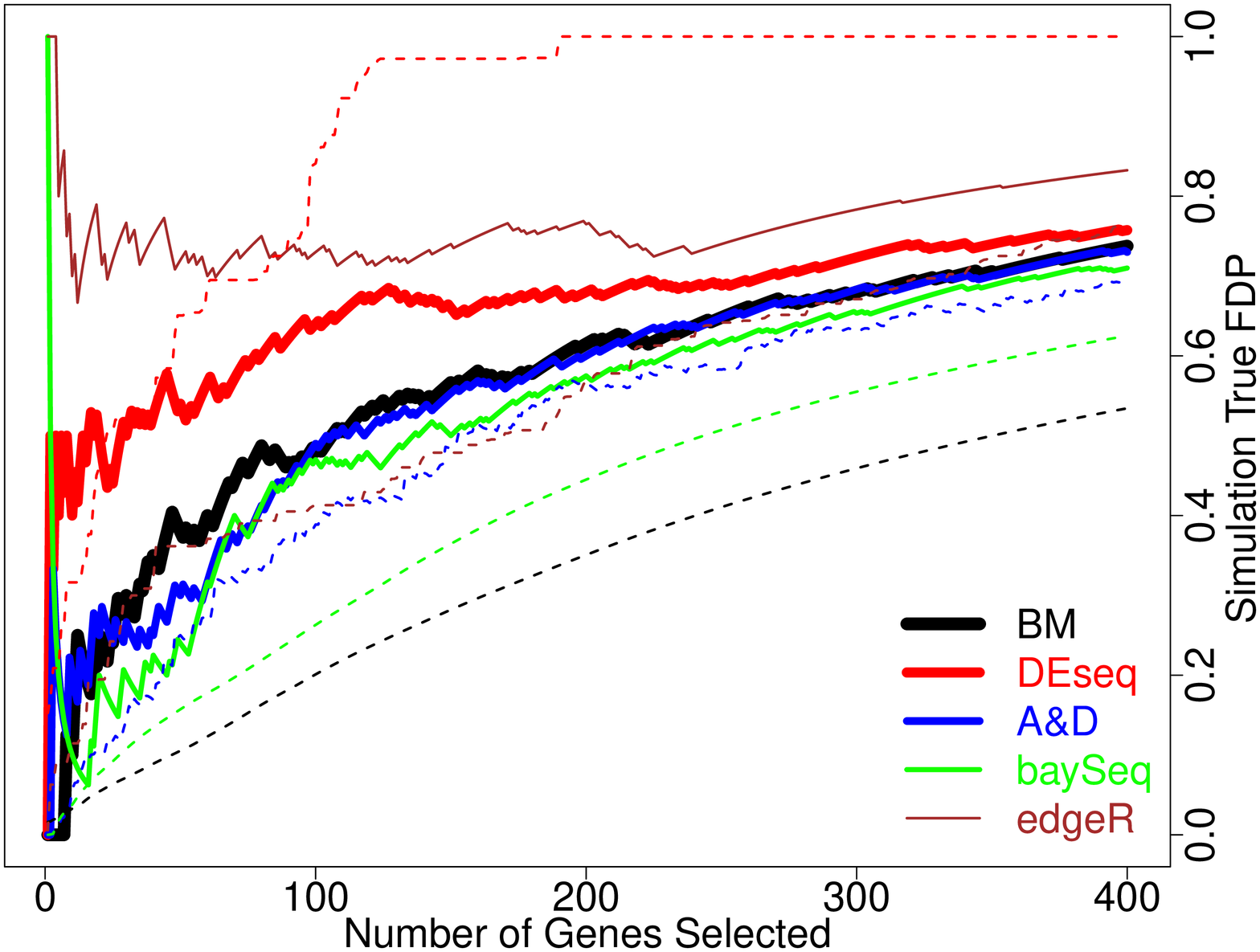}
\end{figure}

\begin{figure}[]
\caption{\label{fig:BMFvsDESeq_combo_Csimulationalpha_exp_v_50_p1nonrandomfavoringDESeq} 
\textsc{\small Comparison of ROC and FDR curves of different procedures under scenario 3 
($\alpha_i$ is a function of $\mu_i$ and $\alpha_i$ happens to be less than $0.07$, 
$\tilde{\pi_1}=0.0108$ with simulation true $\pi_1=0.024$). The numbers in parentheses correspond to the Area Under the curves (AUC).
  }}
\includegraphics[width=2.75in,height=3.25in,angle=270]{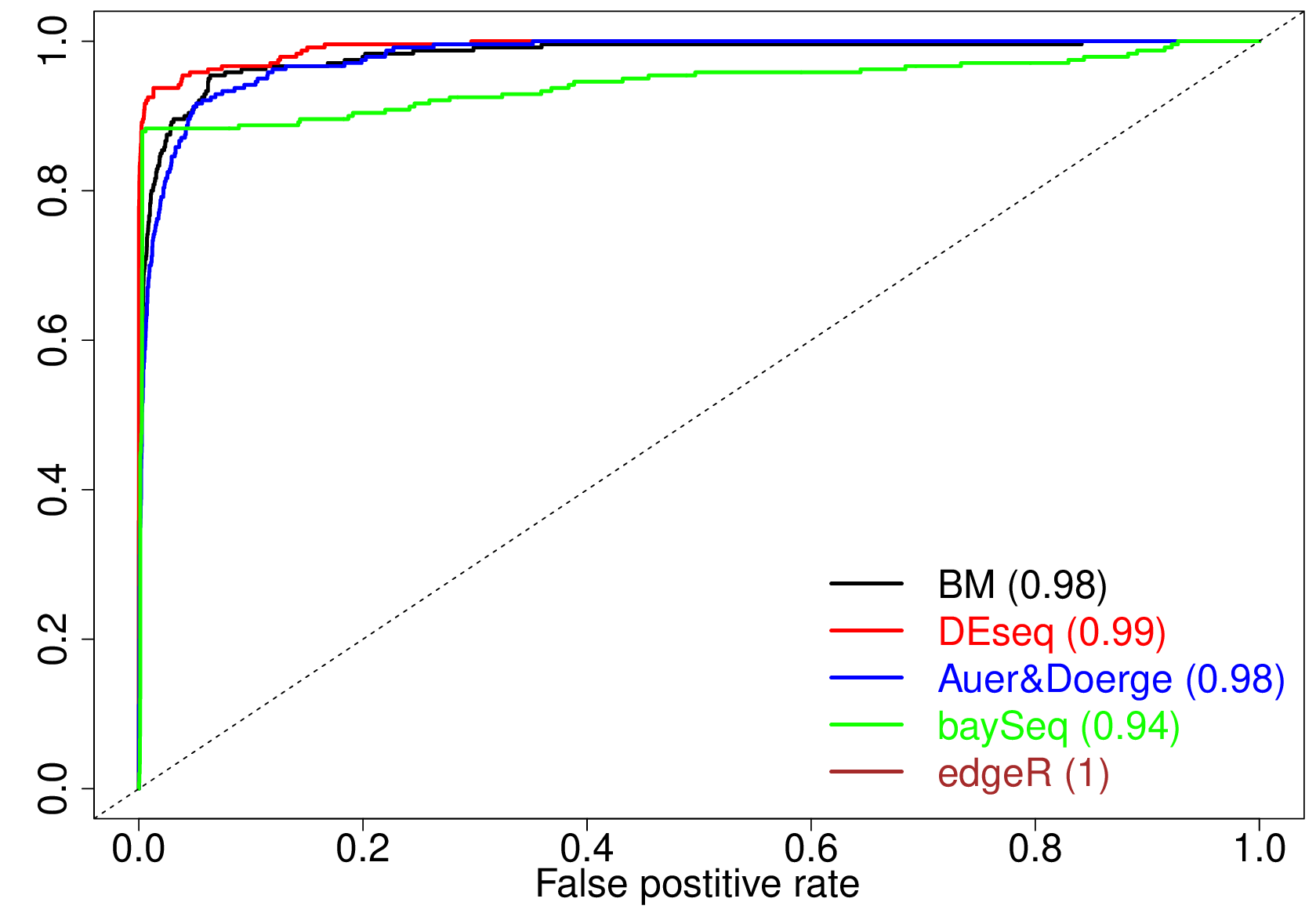}
\includegraphics[width=2.75in,height=3.25in,angle=270]{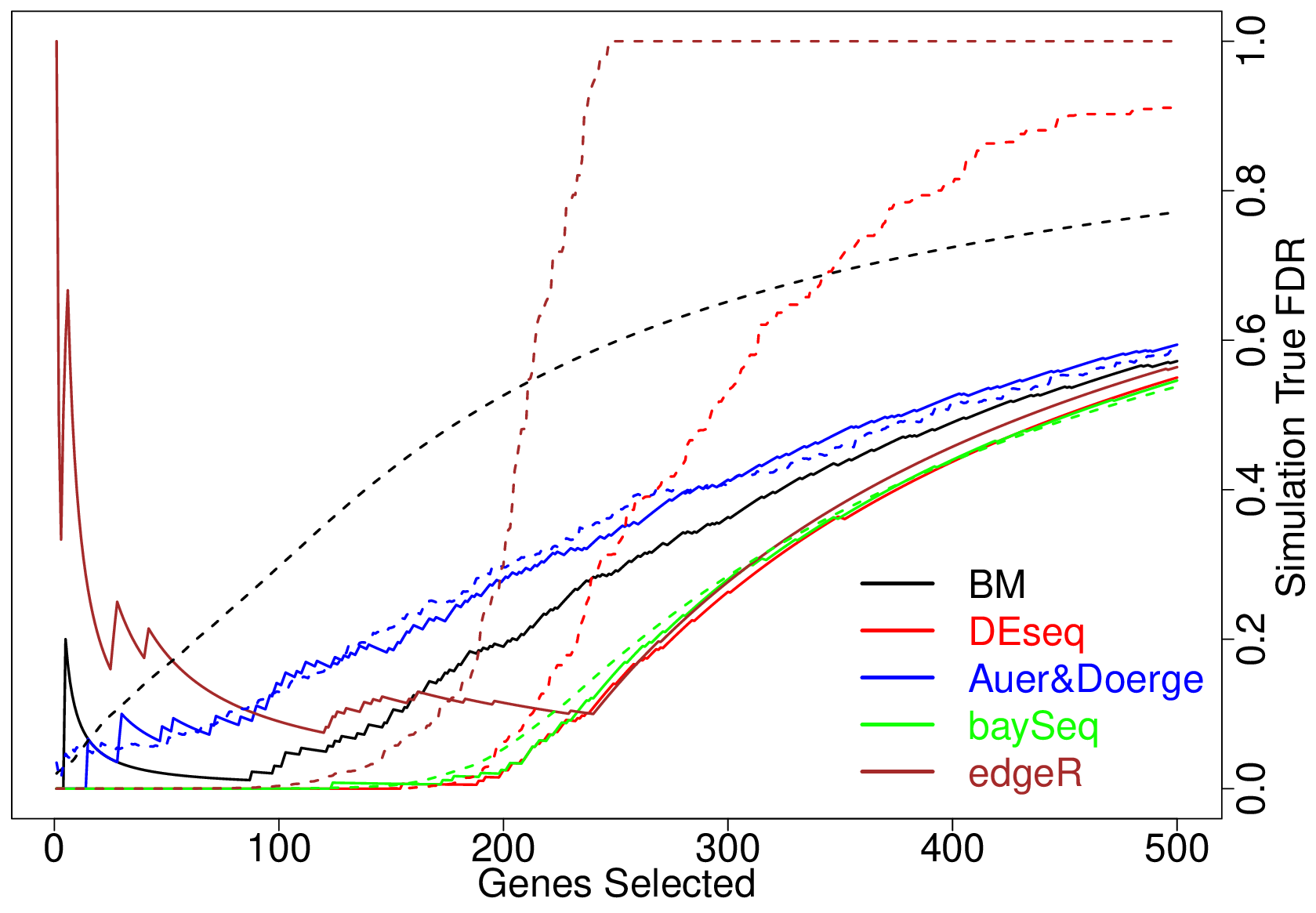}
\end{figure}


Figure  \ref{fig:BMFvsDESeq_combo_Csimulationalpha_exp_v_50_p1nonrandom} compares the results from the Bayes model (BM) with those of DESeq, A\&D, edgeR and baySeq under Study 2. In terms of the ROC curve, all the procedures have a similar behavior with the exception of baySeq
which offers a substantially lower tradeoff between the true positive and false positive proportions, which is expected in this simulation scenario. In terms of FDP (right panel), we observe that BM is competitive with A\&D and baySeq and they all perform better that DESeq in the range of genes that are differentially expressed.


Finally, Figure \ref{fig:BMFvsDESeq_combo_Csimulationalpha_exp_v_50_p1nonrandomfavoringDESeq} summarizes the results on Study 3. Even though the simulations were set to favor DESeq, we observe that the proposed Bayes model remains quite competitive to other alternatives in terms of ROC curves (left panel) and false discoveries (right panel), particularly within range of genes differentially expressed, where the true FDP does not exceed 20\% for 200 selected genes.



\section{Application}\label{sec:application}
\begin{figure}[t]
\caption{\label{figure:realdatafigs}\textsc{\small For the RNA-Seq data described in Section \ref{sec:application}, the left panel shows the posterior expected FDP as a function of the number of genes declared DEG. The horizontal line marks FDP=0.05, yielding a selection of 362 genes.  The right panel shows $\log_2 (\hat{\mu}_{2i}/\hat{\mu}_{1i})$ vs the posterior probability of $H_{1i}$, where $\hat{\mu}_{1i}$ and $\hat{\mu}_{2i}$ are the estimated mean counts for the female and male flies respectively. The points representing flagged as DEG are above the discontinuous horizontal line. }}
\includegraphics[width=2.75in,height=3.25in,angle=270]{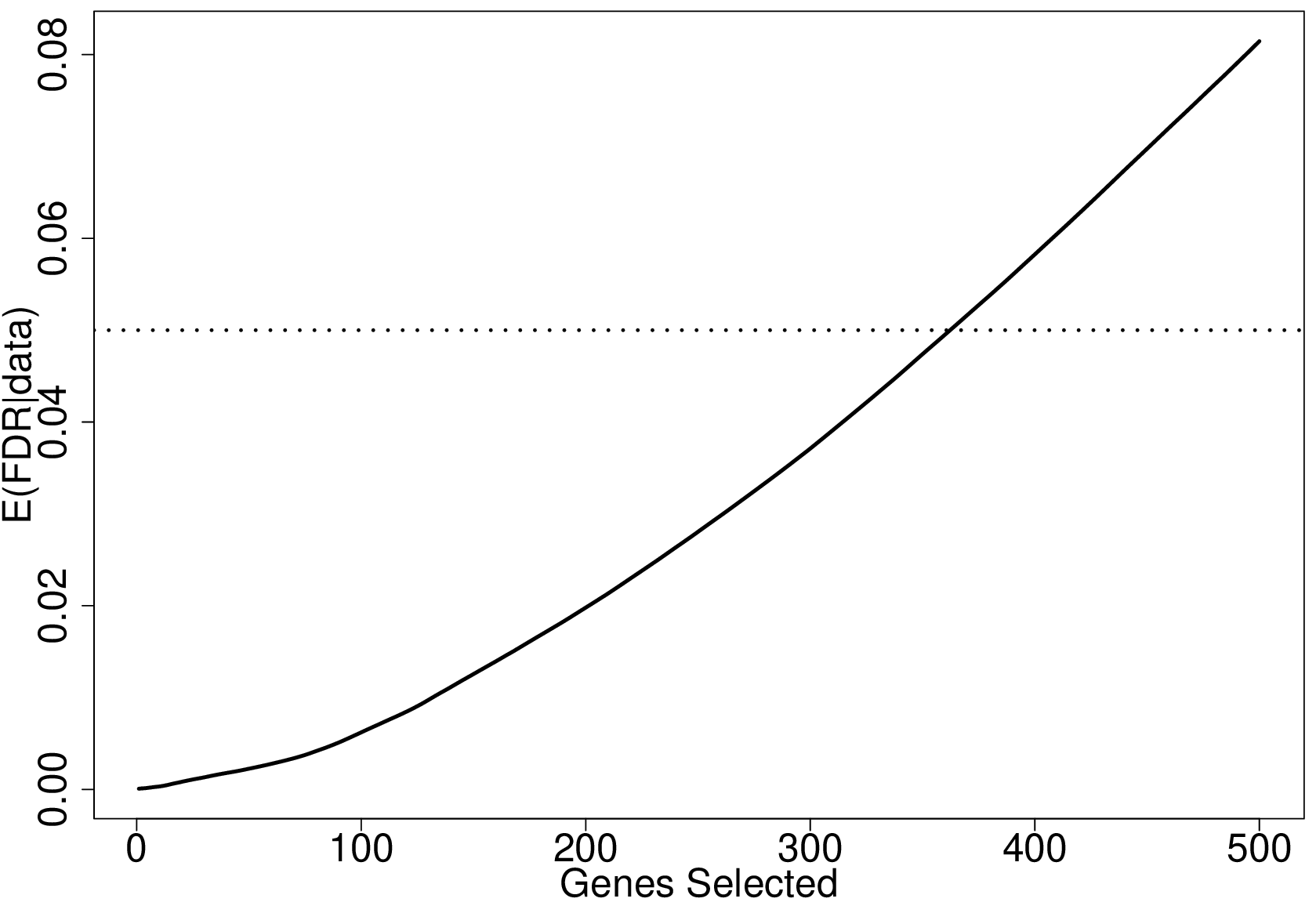}
\includegraphics[width=2.75in,height=3.25in,angle=270]{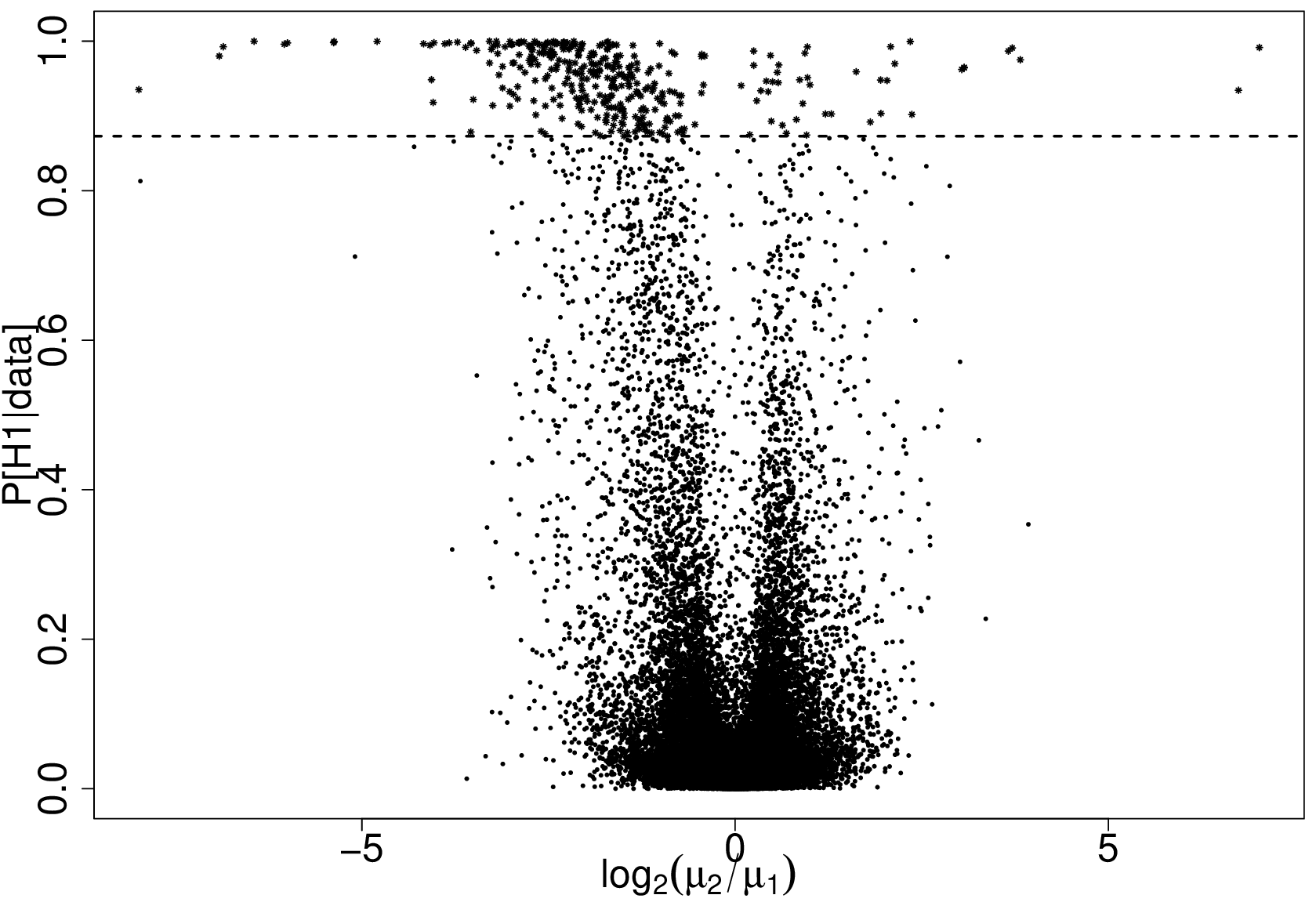}
\end{figure}
In order to illustrate the methodology, e compare the  expression of brain cells in females vs. male flies ({\it Drosophila Melanogaster}) in terms of exons, that, for the sake of the example, we will indistinctively call genes. In the experiment, six bottles of flies were used; $J_1=3$ bottles of female flies and $J_2=3$ of male flies. For every bottle we extract two samples of fly brains  and sequence them obtaining two technical  replications per bottle.  For every exon we consider one  count per bottle  by summing the counts over the two technical replications. Initially, 60277 genes were sequenced, and were pre-processed by the experimenter (removing genes with total sum of read counts less than or equal to 5) resulting in 48944 genes. 
We estimated the abundances according to equation \eqref{libsize} yielding
$\hat{s}_1=0.9378078$, 
$\hat{s}_2=0.9101029$, 
$\hat{s}_3=0.8255321$,
$\hat{s}_4=1.1126044$,
$\hat{s}_5=1.1928032$, and 
$\hat{s}_6=1.1221106$.

We tested the hypotheses in \eqref{twocompetingmodel}, with hyperparameter values set as described in Section \ref{sec:testing} yielding $u_0= 0.437, u_1= 0.411, v_0= 1.568$ and $v_1= 1.640$.

We also estimated the means by
$\hat{\mu}_{i1}=(1/J_1) \sum_{j=1}^{J_1}  k_{ij}/s_j$ and 
$\hat{\mu}_{i2}=(1/J_2) \sum_{j=J_1+1}^{J_1+J_2} k_{ij}/s_j$.
The estimated proportion of exons differentially expressed is $\tilde{\pi}_1=0.04$.
The left panel in Figure \ref{figure:realdatafigs} shows the posterior expected FDP as a function
of the number of selected genes.  Considering an $E(FDP\mid data)$=0.05 we declare 362 DEG. 
These genes have posterior probabilities $H_{1i}$ 
greater than $0.78$.   In the left panel we show a volcano plot comparing the $\log_2 (\hat{\mu}_{2i}/\hat{\mu}_{1i})$ with the posterior probability of being a DEG, where $\hat{\mu}_{1i}$ are the males flies and $\hat{\mu}_{2i}$ are the females flies. We observe that most DEG are overexpressed in the female flies. 

We also analyze the data using DESeq, baySeq, edgeR and A\&D, remember that our approach used the same library sizes as DESeq. To implement A\&D we also used these library sizes.  
For baySeq and edge R the library sizes were estimated 
by the R function \texttt{getLibsizes} in the \texttt{baySeq} package, and
by the R function \texttt{calcNormFactors} in the \texttt{edgeR} package, respectively.

\begin{figure}[!htb]
\caption{\label{figure:venn}\textsc{\small 
Venn diagrams of the top 362 genes selected by BM, DESeq, baySeq (top left)
by BM edgeR and A\&D (bottom left)
and of the DEG reported by these methods when requiring a FDR of 0.05 (right).
  }}
$\begin{array}{rl}
\includegraphics[height=3in,angle=270]{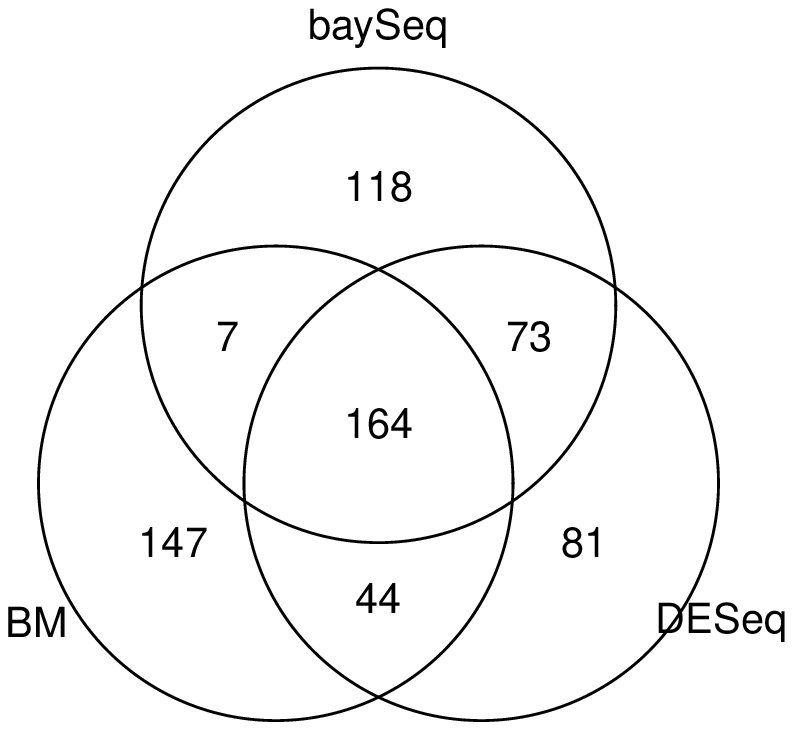}&
\includegraphics[height=3in,angle=270]{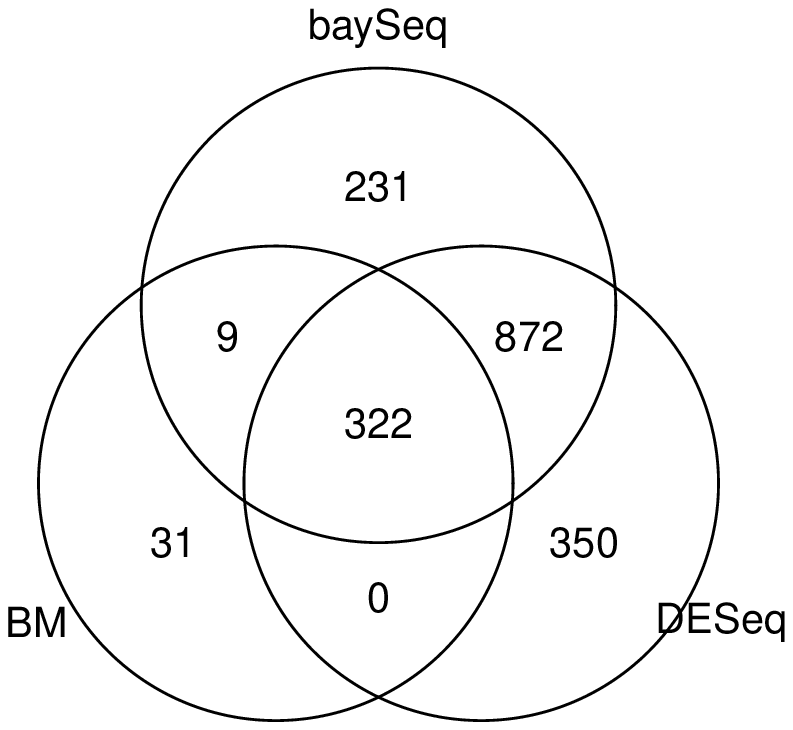}\\
\includegraphics[height=3in,angle=270]{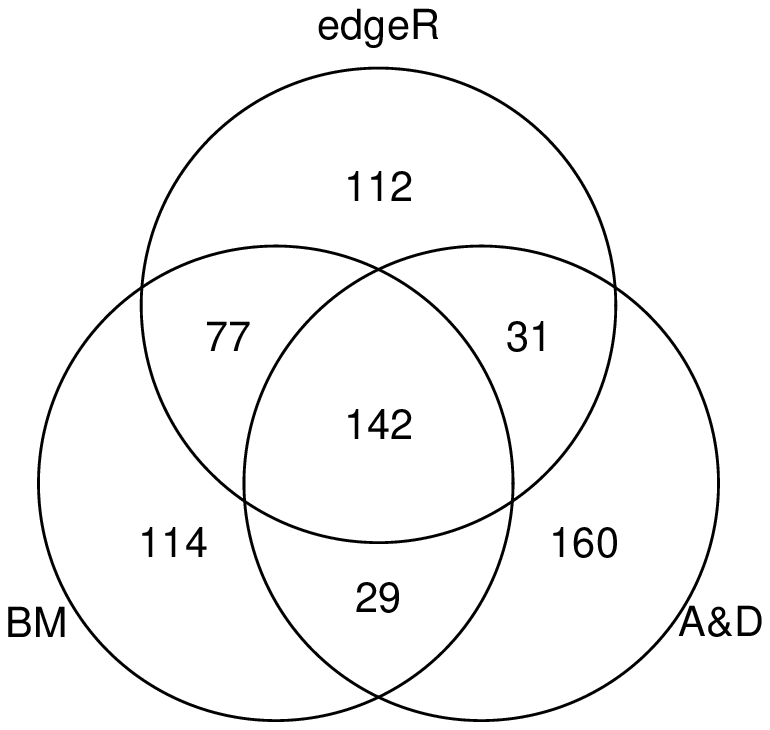}&
\includegraphics[height=3in,angle=270]{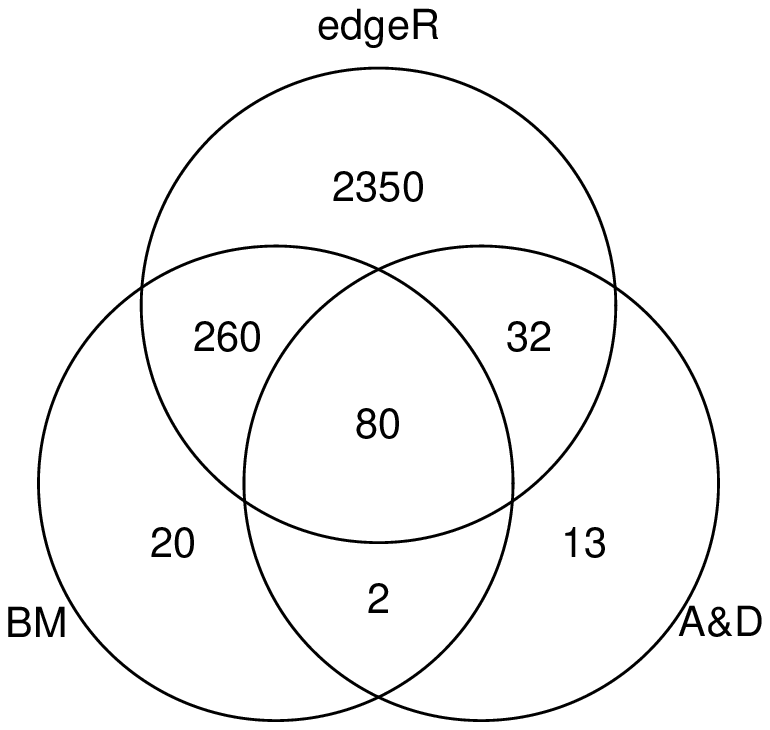}
\end{array}$
\end{figure}

For comparison we consider the top 362 selected by these procedures.
The reported FDPs are 
0.05,
$1.589095\times 10^{-6}$,
$8.445714\times 10^{-6}$,
$4.086\times 10^{-9}$,
and
$7.504545\times 10^{-2}$
 for the BM, DESeq, baySeq, edgeR and A\&D, respectively.  Two Venn diagrams of 
these genes are shown in the left panels  of Figure \ref{figure:venn}.
We also fixed the estimated FDR equal to 0.05 for the five methods
BM, DESeq and baySeq, endR and A\&D report 362, 1544, 1434,
2722 and 127 DEG, respectively. 
The Venn diagrams for these genes are shown in the right panels of Figure \ref{figure:venn}.

\section{Discussion}\label{sec:disc}

In this paper, we first proposed a new estimator of the overdispersion parameter of the negative binomial distribution, obtained via maximization of the marginal likelihood of a conjugate Bayesian model. We showed, using simulation results, that the marginalized MLE, outperforms the standard MLE and the quasilikelihood estimator in \cite{Robinson:Smyth:2008} in terms of MSE, offering a more accurate and stable estimator. The results are particularly promising in small sample scenarios. We then extended the idea behind this estimator for the analysis of RNA-Seq data when only a few counts per gene are available.  Specifically, we used a Bayesian framework to develop a hypothesis test to detect DEG. Our approach, which does not assume any functional relation between the overdispersion parameter and the mean (a standard assumption of other approaches such as DESeq), implicitly considers the marginal distribution of the model in Section \ref{sec:overdispersion}, and therefore, enjoys the same stability properties as the marginal MLE.

Because the inference is carried out gene by gene, our method enjoys of a borrowing-of-strength effect across genes at the level of the prior distribution of the overdispersion parameter $\alpha$.  Furthermore, the Bayesian framework we considered here is flexible and a number of extensions are possible. For instance, although the chosen prior is a gamma distribution with parameters depending on the initial estimates of $\alpha_i$, we could instead assume  that all $\alpha_i$'s come from an underlying prior distribution and make inference considering all genes simultaneously.  Such extensions are not explored in this paper, as a different computing approach would be required.
Unlike other methods, our model imposes fewer assumptions about the number of counts, library sizes and mean-variance relationship. For instance, DESeq assumes that genes with similar mean counts have similar variances. Despite of the fewer assumptions, we showed via simulation studies that our approach remains competitive when compared to DESeq, baySeq, A\&D and edgeR and sometimes performs better than some of the alternatives (in terms of ROC curves and proportion of false discoveries), even in situations designed to favor the competitors.

The proposed model flags genes with the highest Bayes factors as DEG. To accomplish this, we do not require  \textit{a priori} specification of the proportion $\pi_1$ of DEG.  However,  in order to estimate both the posterior probability of a gene being a DEG and  the FDP the method does requires a value for $\pi_1$. To estimate $\pi_1$ we follow the approach by \cite{wen2013robust}, which has proven to produce very accurate estimates of the true FDP in a number of scenarios and perform better than the FDP reported by other methods in a number of cases.

Finally, the simple Bayesian hierarchical structure of our method make the model easy to interpret and implement to analyze real data sets using standard MCMC techniques. The Gibbs sampler algorithm we include in the Appendix compute the Bayes factors of interest within a few minutes and allows for modifications and extensions to further tailor to the dataset and question of interest.  
Software is available from the authors upon request.  

\section*{Acknowledgments}
The authors would like to thank Dr. Lauren McIntyre from the University of Florida for facilitating the data set used in this work and for her insightful comments and suggestions.

\bibliographystyle{apalike}
\bibliography{bibliography_NB}

\appendix
\section{Estimating $P(H_{1i})$ for the unequal abundance case}\label{app:gibbs}
We omit the gene index $i$ of the notation.
We can approximate the Bayes factor in favor of model $H_1$ with the harmonic mean estimator \citep{newton1994approximate}, 
\begin{equation*}
BF_{10}\approx \sum_{m=1}^M {1/\ell_0(\mu_0^{(m)},\alpha_0^{(m)};\kv)}\Bigg/    \sum_{m=1}^M {1/\ell_1(\mu_1^{(m)},\mu_2^{(m)},\alpha_1^{(m)};\kv)},
\end{equation*}
where
\begin{equation*}
\ell_0(\mu_0,\alpha_0;\kv) =\prod_j \text{NegBin}(k_j\mid \mu_0,\a_0)
\end{equation*}
and
\begin{equation*}
\begin{array}{rl}
\ell_1(\mu_1,\mu_2,\alpha_1;\kv)=& \prod_{j=1}^{J_1} \text{NegBin}(k_j\mid \mu_1,\a_1)\\
&\times\prod_{j=J_1+1}^{J_1+J_2} \text{NegBin}(k_j\mid \mu_2,\a_1)
\end{array}
\end{equation*}
are the likelihoods under $H_0$ and $H_1$ respectively, and $\gamma^{(m)}$ denotes the $m$-th random number generated from
the posterior distribution  of the generic parameter $\gamma$. 
These random numbers  can be obtained running two Gibbs samplers (one under $H_0$ and one under $H_1$),


The full conditional distribution under $H_0$ is
\begin{equation*}
\begin{array}{rl}
p(\mu_0,\a_0 \mid \kv)\propto&
\ell_0(\mu_0,\alpha_0;\kv)\times \text{F}(\mu_0\mid 2 a_\mu, 2 a_\mu/\alpha_0) \times\text{gamma}(\alpha_0\mid u_0,v_0)\\
=&
\prod_j\left[\frac{1}{{\cal B}(k_j+1,\a_0^{-1})(k_j+\a_0^{-1})}
\frac{(\a_0 s_j\mu_0)^{k_j}}{(\a_0 s_j\mu_0+1)^{k_j+1/\a_0}}
\right]\\
\times&\frac{1}{{\cal B}(a_\mu,a_\mu/\a_0)}
\frac{\a_0^{a_\mu}\mu_0^{a_\mu-1}  }{(1+\a_0\mu_0)^{(1+1/\a_0)a_\mu}}
\times \alpha_0^{u_0-1}v_0 \exp(-v_0 \alpha_0)\\
\end{array}
\end{equation*}
and under $H_1$,
\begin{equation*}
\begin{array}{rl}
p(\mu_1,\mu_2,\a_1  \mid\kv)\propto&
\ell_1(\mu_1,\mu2,\alpha_1;\kv)\times \text{F}(\mu_1\mid 2 a_\mu, 2 a_\mu/\alpha_1) 
\times \text{F}(\mu_2\mid 2 a_\mu, 2 a_\mu/\alpha_1)
\times\text{gamma}(\alpha_1\mid u_1,v_1)\\
\times&\prod_{j=1}^{J_1+J_2}\left[\frac{1}{{\cal B}(k_j+1,\a_1^{-1})(k_j+\a_1^{-1})}\right]\\
\times&\left[
\prod_{j=1}^{J_1}
\frac{(\a_1 s_j\mu_1)^{k_j}}{(\a_1 s_j\mu_1+1)^{k_j+1/\a_1}}
\right]\times
\left[ \prod_{j=J_1+1}^{J_1+J_2}
\frac{(\a_1 s_j\mu_2)^{k_j}}{(\a_1 s_j\mu_2+1)^{k_j+1/\a_1}}
\right]\\
\times&\frac{1}{{\cal B}^2(a_\mu,a_\mu/\a_1)}
\frac{\a_1^{a_\mu}\mu_1^{a_\mu-1}  }{(1+\a_1\mu_0)^{(1+1/\a_1)a_\mu}}
\times
\frac{\a_1^{a_\mu}\mu_2^{a_\mu-1}  }{(1+\a_1\mu_0)^{(1+1/\a_1)a_\mu}}
\times \alpha_1^{u_1-1} v_1 \exp(-v_1 \alpha_1)\\
\end{array}
\end{equation*}

The complete conditional distributions for the Gibbs sampler (we denote as $all$ all the conditioning parameters) is given by
\begin{equation*}
p(\mu_i\mid all)\propto \frac{\mu^{\sum_j k_j+a_\mu-1}    }{\prod_j (1+\a s_j\mu_i)^{k_j+1/\a}}
\frac{1}{(1+\a\mu_i)^{a_\mu+a_\mu/\a}}
\end{equation*}
where $k_{\cdot}=\sum_{j} k_j$, the indices in this sum and the product in the equation above depend on  which $\mu$ is being generated.  For $\mu_0,\mu_1$ and $\mu_2$, the index $j$ runs over $1,\dots,J_1+J2$, $1,\dots,J_1$ and $J_1+1,\dots,J_1+J_2$ respectively. Also 
 $\a=\a_0$ when generating $\mu_0$ while $\a=\a_1$  when generating $\mu_1$ and $\mu_2$.  A good starting point for the Gibbs sampler is the average of $ k_j/s_j$. 
 
The Gibbs sampler can be implemented in two different ways. The first one is through a Metropolis-Hastings (MH) random walk with proposal distribution $N(\mu^t,\sigma_\mu^2)$, where $\sigma_\mu^2$ is set equal to the sample variance of $k_j/s_j$. 
The second, generating $\mu^{t+1}$ by a MH algorithm  that uses as candidate the posterior distribution of $\mu$, if all the abundances were 1. Such distribution is an F- distribution. The MH algorithm 
is: 

\begin{enumerate}
\item Generate $\mu^\prime\sim \text{F}\left(2(a_\mu+\sum_j k_j),2(J+a_\mu)/\a\right)$
\item Set $\mu^c=\mu^\prime\times (a_\mu+\sum_j k_j)/(J+a_\mu)$
\item Set $\mu^{t+1}=\mu^c$ with probability $\min\{1,\exp(\beta)\}$ with
\begin{equation*}
\beta = \sum_j(k_{j}+1/\a)\log\left[ \frac{1+\a s_j\mu^t}{1+\a s_j\mu^c}\right]+(\sum_j k_j+J/\a)\log\left[\frac{1+\a\mu^c}{1+\a\mu^t} \right]
\end{equation*}
Here $J$ is $J_1+J_2$, $J_1$ or $J_2$ when generating $\mu_0,\mu_1$ or $\mu_2$ respectively. 
\end{enumerate}

We simulate $\a_0$ and $\a_1$ via MH. The $\log$ of the complete posterior distributions, up to additive constants not depending on $\a$, are given by 
\begin{equation*}
\begin{array}{rl}
\log p(\a_0\mid all)=&
-\sum_{j=1}^{J_1+J_2}\left[
\log {\cal B}(k_{j}+1,\a_0^{-1})+\log(k_{j}+\a_0^{-1})\right]\\
&+\left[a_\mu+\sum_{j=1}^{J_1+J_2} k_j \right]\log \a_0 \\
&-\sum_{j=1}^{J_1+J_2}(k_j+\a_0^{-1}) \log(\a_0\mu_0 s_j+1)\\
&-\log {\cal B}(a_\mu,a_\mu/\a_0)-a_\mu(1+\a_0^{-1})\log(1+\a_0\mu_0) \\
&+(u_0-1)\log(\alpha_0)-v_0 \a_0,
\end{array}
\end{equation*}

\begin{equation*}
\begin{array}{rl}
\log p(\a_1\mid all)=&
-\sum_{j=1}^{J_1+J_2}\left[
\log {\cal B}(k_{j}+1,\a_1^{-1})+\log(k_{j}+\a_1^{-1})\right]\\
&+\left[2 a_\mu+\sum_{j=1}^{J_1+J_2} k_j \right]\log \a_1 \\
&-\sum_{j=1}^{J_1}(k_j+\a_1^{-1}) \log(\a_1\mu_1 s_j+1)-\sum_{j=J_1+1}^{J_1+J_2}(k_j+\a_1^{-1}) \log(\a_1\mu_2 s_j+1)\\
&-2 \log {\cal B}(a_\mu,a_\mu/\a_1)-a_\mu(1+\a_1^{-1})\left[\log(1+\a_1\mu_1)+\log(1+\a_1\mu_2)\right] \\
&+(u_1-1)\log(\alpha_1)-v_1 \a_1.
\end{array}
\end{equation*}

\end{document}